%% file: HDRKC25.tex
\documentclass[12pt]{article}
\usepackage{amsfonts,amssymb,amsmath,amscd,theorem,oldgerm,epsfig}
\usepackage[mathscr]{eucal}
\usepackage{mathrsfs}

\input diagram.tex 

\newtheorem{lemma}{Lemma}[section]

\newtheorem{theorem}[lemma]{Theorem}

\newcommand{\EProof}{\hfill\blacksquare}

\def\ome {\omega}

\def\XI{\xi}
\def\XII{\eta}
\def\CHI{\chi} 
 
\def\lto{\longrightarrow}

\def\iso{\backsimeq}

\def\R{{\mathbb R}}
\def\Q{{\mathbb Q}}

\def\Z{{\mathbb Z}}
\def\C{{\mathbb C}}

\def\Fp{\mathbb{F}_{p}}
\def\FFp{\overline{\mathbb{F}}_{p}}
\def\Fq{{\mathbb{F}}_{q}}

\def\lam{\lambda}
\def\Lam{\Lambda}
\def\Lm{\Lambda^*}
\def\Lmm {\Lambda^{*}_{p}}
\def\LmmV{\overline{\Lambda^{*}}_{p}}

\def\GL{\mathrm{GL}}
\def\PGL{\mathrm{PGL}}

\def\G{\mathrm G}
\def\GG{Sp(2n,\mathbb{F}_{p})}
\def\GGV{Sp(2n,\overline{\mathbb{F}}_{p})}
\def\GrA{\langle A \rangle}
\def\CA{{\cal C}_{A}}
\def\CAV{\overline{{\cal C}}_{A}}

\def\GrA{\langle A \rangle}
\def\CA{{\cal C}_{A}}
\def\CAV{\overline{{\cal C}}_{A}}

\def\FT{{\cal F}(\T)}

\def\T{{\bf T}}
\def\A{\cal A}

\def\Irr{\mathrm{Irr}({\cal A} _\hbar)}
\def\Ad{ {\cal A} _\hbar}

\def\hb{\hbar}
\def\h{ \hbar }

\def\H{{\cal H}}
\def\Hh{{\cal H} _\h}

\def\rhoh{\rho _{_\h}}
\def\pih{\pi i {\h}}
\def\Pih{\pi_{_\hbar}}

\def\Y{Y_{0}}
\def\YY{Y}
\def\YYV{\overline{Y}}
\def\TTV{\overline{T}}

\def\XI{\xi}

\def\XII{\eta}

\def\i_XI{i_{_{\XI}}}
\def\iXII{i_{_{\XII}}}
\def\p_XI{p_{_{\XI}}}
\def\pXII{p_{_{\XII}}}
\def\S0{S}

\def\SF{\mathcal{F}}
\def\SG{\mathcal{G}}
\def\SL{\mathscr L}

\def\Tr{\mathrm{Tr}}
\def\Av{\mathrm{\bf Av}}
\def\ev{{\mathrm{e.v}}}

\def\End{\mathrm{End}}

\def\Db{{\cal D}^{b}_{\mathrm{c,w}}}
\def\Fr{\mathrm{Fr}}

\def\mchi{m(\chi)}

\def\r{\;}

\def\half{{1 \over 2}}

\begin{document}

\title{\bf \small THE HIGHER-DIMENSIONAL RUDNICK-KURLBERG CONJECTURE}
\author{\it \small SHAMGAR GUREVICH AND RONNY HADANI}

\date{preliminary version}

\maketitle

\bigskip

\begin{abstract}
In this paper we give a proof of the {\it Hecke quantum unique
ergodicity conjecture} for the multidimensional Berry-Hannay model.
A model of quantum mechanics on the 2n-dimensional torus.
This result generalizes the proof of the {\it Rudnick-Kurlberg Conjecture} given in \cite{GH3} 
for the 2-dimensional torus.
\end{abstract}

\bigskip

\tableofcontents

\setcounter{section}{-1}
\section{Introduction}
\subsection{Berry-Hannay model} In the paper ``{\it Quantization of linear maps on the torus -
Fresnel diffraction by a periodic grating}'' , published in 1980
(see \cite{BH}), the physicists M.V. Berry and J. Hannay explore a
model for quantum mechanics on the 2-dimensional symplectic torus
$(\T,\ome)$. Berry and Hannay suggested to quantize simultaneously
the functions on the torus and the linear symplectic group
$\G=SL(2,\Z)$.

\subsection{Quantum chaos} One of the motivations for considering the Berry-Hannay model was to study the phenomenon
of quantum chaos in this
model (see \cite{R2} for a survey).\\

Consider a classical ergodic mechanical system. In what sense the ergodicity property is reflected in
the corresponding quantum system ?. Or phrase it differently, is there a meaningful notion of {\it
Quantum Ergodicity} ?. This is a fundamental meta-question in the area of quantum chaos.\\

\subsection{Hecke quantum unique ergodicity} For the specific case of the 2-dimensional Berry-Hannay model, the quantum ergodicity question was
addressed in a paper by  Rudnick and Kurlberg \cite{KR}. In that
paper they formulated a rigorous definition of quantum ergodicity.
We call this notion {\it Hecke Quantum Unique Ergodicity}. A focal
step in their work was to introduce a group of hidden symmetries
they named the {\it Hecke} group. The statement of {\it Hecke
Quantum Unique Ergodicity} concerns the semi-classical convergence
of certain matrix coefficients which are defined in terms of the
{\it Hecke} group. In their paper they proved a bound on the rate
of convergence. In Z. Rudnick's lectures at MSRI, Berkeley 1999
\cite{R1} and ECM, Barcelona 2000 \cite{R2} he conjectured that a stronger bound should hold true.\\

In \cite{GH1} we have constructed the 2-dimensional Berry-Hannay model and in \cite{GH3}  we have proved the 2-dimensional Rudnick-Kurlberg rate conjecture.
We established the correct bound, on the semi-classical
asymptotic of the {\it Hecke} matrix coefficients, as stated in Rudnick's lectures
\cite{R1, R2}.\\

\subsection{Motivation} Our motivation in this paper is to generalize the Rudnick-Kurlberg {\it Hecke theory} to the Multidimensional Berry-Hannay Model (see \cite{GH2}).
More specifically we study the higher-dimensional generalization of the Rudnick-Kurlberg Conjecture \cite{GH3} about Hecke quantum unique ergodicity.\\

\subsection{Results} In \cite{GH2} we have constructed the {\it 2n-dimensional Berry-Hannay model}.  We have quantized simultaneously
the functions on the 2n-dimensional torus and the linear symplectic group
$\G=Sp(2n,\Z)$. In this paper we develop the {\it Hecke theory} for this model and we manifest the algebro-geometric tools 
to deal with quantitative quantum mechanical statistical questions arising in the higher-dimensional Berry-Hannay model. 
As an application we prove the higher-dimensional {\it Hecke Quantum Unique Ergodicity Conjecture} for ergodic element $A \in Sp(2n,\Z)$
which has an irreducible characteristic polynomial over $\Q$.

\subsection*{Acknowledgments}
We thank warmly our Ph.D. adviser Joseph Bernstein for his interest
and guidance along this project. We thank very much P. Kurlberg,
Z. Rudnick and Dubi Kelmer who discussed with us their papers and explained their
results. We thank P. Sarnak for the discussion on our work. 
We thank our friends D. Gaitsgory and L. Ramero for the discussions on
$\ell$-adic sheaves, $\ell$-adic cohomologies and Deligne's
letter. This paper was written while we were visitors at the Max Planck Institute and Bonn university in the Summer 2004.
We thank Prof. W. Mueller, Prof. P. Moree and the nice people at MPI for the invitation. 

\section{Classical Torus}
Let $(\T,\ome)$ be the 2n-dimensional symplectic torus. 
Together with its linear
symplectomorphisms $\G \simeq Sp(2n,\Z)$ it serves as a simple model of
classical mechanics (a compact version of the phase space of the
harmonic oscillator). More precisely, Let $\T=V/\Lam$, where $V$ is
a 2n-dimensional real vector space, $V \simeq \R^{2n}$ and $\Lam$ is
a full lattice in $V$, $\Lam \simeq \Z^{2n}$. The symplectic form on
$\T$ we obtain by taking a non-degenerate symplectic form on $V$,

 $$\ome:V \times V \lto \R$$

We require $\ome$ to be integral, namely $\ome : \Lam \times \Lam \lto \Z$ and normalized such that Vol$(\T) =1$.\\

Let $Sp(V,\ome)$ be the group of linear symplectomorphisms, $Sp(V,\ome) \simeq Sp(2n,\R)$. Consider the subgroup $\G
\subset Sp(V,\ome)$ of elements which preserve the lattice $\Lam$, i.e $\G (\Lam) \subseteq \Lam$. Then
$\G \simeq Sp(2n,\Z)$. The subgroup $\G$ is exactly the group of linear symplectomorphisms of $\T$.\\

We denote by $\Lm \subseteq V^*$ the dual lattice, $\Lm  = \{ \xi \in V^* | \r\r \xi (\Lam) \subset \Z \} $. The
lattice $\Lm$ is identified with the lattice of characters of $\T$ by the following map:

\begin{equation}\label{ident}
  \xi \in \Lm \longmapsto e^{2 \pi i <\xi, \cdot>} \in \r \T^{*},
\end{equation}
where $\T^{*} = Hom(\T,\C^{\times})$.
\subsection{Classical mechanical system}

We consider a very simple discrete mechanical system. An element $A \in \G$, that has no roots of unity as eigenvalues, generates an
ergodic discrete dynamical system. The {\it ergodic
theorem} asserts that:
\begin{equation}\label{classical ergodicity}
  \lim_{N \rightarrow \infty} \frac{1}{N}\sum_{k=1}^{N} f(A^k
  x)= \int_{\T}f \ome,
\end{equation}
for every $f \in \FT$ and almost every point $x \in \T$. Here
$\FT$ stands for a good class of functions, for example
trigonometric polynomials or smooth functions.\\

\section {Quantization of the Torus}
Quantization is one of the big  mysteries of modern mathematics, indeed it is not clear at all what is the precise
structure which underlies quantization in general. Although physicists have been quantizing for almost a
century, for mathematicians the concept remains all-together unclear. Yet in specific cases there are certain
formal models for quantization which are well justified on the mathematical side. The case of the symplectic torus
is one of these cases. Before we employ the formal model, it is worthwhile to discuss the general phenomenological
principles
of quantization which are surely common for all models.\\

Let us start from a model of classical mechanics, namely a
symplectic manifold, serving as classical phase space. In our case
this manifold is the symplectic torus $\T$. Principally
quantization is a protocol by which one associates to the
classical phase space $\T$ a quantum "phase" space $\H$, where
$\H$ is a Hilbert space. In addition the protocol gives a rule by
which one associates to every classical observable, namely a
function $f \in \FT$ a quantum observable ${\cal O}_{f} : \H \lto
\H$, an operator on the Hilbert space. This rule should send a real
function into a self adjoint operator.\\

To be a little bit more precise, quantization should be considered
not as a single protocol, but as a one parametric family of
protocols, parameterized by $\hb$, the Planck constant. For every
fixed value of the parameter $\hb$ there is a protocol which
associates to $\T$ a Hilbert space $\H_{\hb}$ and for every
function $f \in \FT$ an operator ${\cal O}_{f}^{\hb} : \H_{\hb}
\lto \H_{\hb} $. Again the association rule should send real
functions to self adjoint operators.\\

Accepting the general principles of quantization, one searches for
a formal model by which to quantize, that is a mathematical model
which will manufacture a family of Hilbert spaces $\H_{\hb}$ and
association rules $\FT \rightarrow \End(\H_{\hb})$. In this work we
employ a model of quantization called the {\it Weyl Quantization
model}.

\subsection {The Weyl quantization model}

The Weyl quantization model works as follows. Let $\Ad$ be one
parametric deformation of the algebra $\A$ of trigonometric
polynomials on the torus. This algebra is known in the literature
as the Rieffel torus \cite{Ri}. The algebra $\Ad $ is constructed
by taking the free algebra over $\C$ generated by the following
symbols $\{s(\xi) \r | \r \xi \in \Lm \}$ and quotient out by the
relation $ s(\xi + \eta) = e^{\pih \ome(\xi,\eta)}s(\xi)s(\eta)$.
We point out two facts about this algebra $\Ad$. First when substituting $\hb
= 0$ one gets the group algebra of $\Lm$, which is exactly equal
to the algebra of trigonometric polynomials on the torus. Second
the algebra $\Ad$ contains as a standard basis the lattice $\Lm$:

\begin{equation}\label{basis}
s: \Lm \lto \Ad
\end{equation}

So one can identify the algebras $\Ad \simeq \A$ as vector spaces.
Therefore every function $f \in \A$ can be viewed as an element of
$\Ad$.\\

For a fixed $\hbar$ a representation $\Pih:\Ad \lto \End(\Hh)$
serves as a quantization protocol, namely for every function $f
\in \A$ one has:

\begin{equation}\label{quantprotocol}
 f \in {\A}  \simeq {\Ad}  \longmapsto \Pih (f) \in \End(\Hh)
\end{equation}

An equivalent way of saying this, is that:

\begin{equation}\label{quantprot2}
  f \longmapsto \sum_{\xi \in \Lm} a_{\xi} \Pih (\xi)
\end{equation}

where  $f = \sum\limits_{\xi \in \Lm} a_{\xi} \cdot \xi $ is the Fourier expansion of $f$.\\

To summarize, every family of representations $\Pih : \Ad \lto
\End(\Hh)$ gives us a complete quantization protocol. But now a
serious question rises, namely what representations to pick ? Is
there a correct choice of representations, both on the
mathematical side, but also perhaps on the physical side?. A
possible restriction on the choice is to pick an irreducible
representation. Yet some ambiguity still remains, because there
are several irreducible classes for specific values of $\hbar$.\\

We present here a partial solution to this problem in the case where the parameter $\hbar$ is restricted to take
only rational values \cite{GH2}. Even more particulary, we take $\hbar$ to be of the form $\hbar = \frac{1}{p}$,
where $p$ is an odd prime number. Before any formal discussion, recall that our classical object is the
symplectic torus $\T$ {\bf together} with its linear symplectomorphisms $\G$. We would like to quantize not only
the observables $\A$, but also the symmetries $\G$. Next we are going to construct an equivariant quantization of
$\T$.

\subsection {Equivariant Weyl quantization of the torus} \label{weilquant}

Fix $\hbar = \frac{1}{p}$. We give here a slightly different
presentation of the algebra $\Ad$. Write $\nu =
\frac{p+1}{2}$. Let $\Ad$ be the free $\C$-algebra generated by
the symbols $\{s(\xi) \;|\; \xi \in \Lm \}$ and the relations
$s(\xi + \eta) = e^{2 \pih \nu \ome(\xi,\eta)}s(\xi)s(\eta)$. The
lattice $\Lm$ serves as a standard basis for $\Ad$:

\begin{equation}\label{basis2}
s: \Lm \lto \Ad.
\end{equation}

The group $\G$ acts on the lattice $\Lm$, therefore it acts on
$\Ad$. It is easy to see that $\G$ acts on $\Ad$ by homomorphisms
of the algebra. For an element $B \in \G$, we denote by $f
\longmapsto f^B$ the action of $B$ on an element $f \in \Ad$.\\

An equivariant quantization of the torus is a pair:

\begin{equation}\label{eqvquant1}
\Pih: {\Ad} \lto \End(\Hh) ;
\end{equation}

\begin{equation}\label{eqvquant2}
\rhoh: \G \lto \PGL(\Hh),
\end{equation}

where $\Pih$ is a representation of $\Ad$ and $\rhoh$ is a
projective representation of $\G$. These two should be compatible
in the following manner:

\begin{equation}\label{eqvquant3}
  \rhoh(B) \Pih(f) \rhoh(B)^{-1} = \Pih(f^B),
\end{equation}

for every $B \in \G$, and $f \in \Ad$. Equation (\ref{eqvquant3})  is called the  {\it Egorov identity}.\\

We give here a construction of an equivariant quantization of the
torus.\\

Given a representation $\pi: \Ad \lto \End(\H)$, and an element $B \in \G$, we construct a new representation $\pi^B:\Ad
\lto \End(\H)$:

\begin{equation}\label{actcat}
  \pi^B(f) := \pi(f^B).
\end{equation}

This gives an action of $\G$ on the set $\Irr$ of equivalence classes of irreducible representations. The set $\Irr$ has a
very regular structure, it is a principal homogeneous space over $\T$. Moreover every irreducible representation
of $\Ad$ is finite dimensional and of dimension $p^n$. The following theorem plays a central role in the
construction.

\begin{theorem}[Canonical invariant representation \cite{GH2}]\label{GH} Let $\hbar = \frac{1}{p}$. There exists a {\bf unique} $($up to
    isomorphism$)$ irreducible representation $(\Pih,\Hh)$ of $\Ad$ for which its equivalence class is fixed by
$\G$.
\end{theorem}

Let $(\Pih,\Hh)$ be a representative of the fixed irreducible
equivalence class. This means that for every $B \in \G$:

\begin{equation}\label{iso}
\Pih^{B} \simeq \Pih.
\end{equation}

Hence for every element $B \in \G$ there exists an
operator $\rhoh(B):\Hh\lto\Hh$ which realizes the isomorphism
(\ref{iso}). The collection $\{\rhoh(B) : B \in \G \}$ constitutes
a projective representation:

\begin{equation}\label{projrep}
  \rhoh:\G \lto \PGL(\Hh)
\end{equation}

Equations (\ref{actcat}), (\ref{iso}) also implies the  Egorov identity (\ref{eqvquant3}).\\

We want to understand if the projective representation (\ref{projrep}) of $\G$ can be lifted (linearized) into a honest representation. 
The next theorem claims the existence of a canonical linearization.

\begin{theorem}[Canonical linearization]\label{GH2} Let $\hbar =
\frac{1}{p}, \r p>2$. there exists a {\bf unique} linearization, which
we denote also by $\rhoh$,

\begin{equation}\label{canstrict}
\rhoh:\G \lto \GL(\Hh),
\end{equation}

characterized by the property that it is factorized through the quotient group $\GG$:
\[
\qtriangle[\G`\GG`\GL(\Hh);`\rhoh`\bar{\rho}_\h]
\]
\end{theorem}

{\bf Summary}. Theorem \ref{GH} claims the existence of a unique
invariant representation of $\Ad$, for every $\hbar = \frac{1}{p}
\r, p>2$. This gives a canonical equivariant quantization
$(\Pih,\rhoh,\Hh)$. Moreover, by Theorem \ref{GH2}, the
projective representation $\rhoh$ can be linearized in a canonical
way to give an honest representation of $\G$ which factories
through $\GG$. Altogether this gives a pair:

\begin{equation}\label{eqvquant1b}
\Pih: {\Ad} \lto \End(\Hh)
\end{equation}

\begin{equation}\label{eqvquant2b}
\rhoh: \GG \lto \PGL(\Hh)
\end{equation}

satisfying the following compatibility condition (Egorov identity):

\begin{equation}\label{eqvquant3b}
  \rhoh(B) \Pih(f) \rhoh(B)^{-1} = \Pih(f^B).
\end{equation}

For every $B \in \GG$, $f \in \Ad$. Here the notation $\Pih(f^B)$
means, to take any pre-image $\bar{B} \in \G$ of $B \in \GG$ and
act by it on $f$, $\Pih(f^{\bar{B}})$ does not depend on the
choice of $\bar{B}$. For what follows, we denote $\bar{\rho}_\h$ by
$\rhoh$, and consider $\GG$ to be the default domain.

\subsection{Quantum mechanical system}

Let $(\Pih,\rhoh,\Hh)$ be the canonical equivariant quantization.
Let $A$ be our fixed ergodic element, considered as an element
of $\GG$ . The element $A$ generates a quantum dynamical system.
For every (pure) quantum state $v \in S(\Hh) = \{ v \in \Hh
:\|v\|=1\}$:

\begin{equation}\label{Qsystem}
  v \longmapsto v^A = \rhoh(A)v
\end{equation}

\section{Hecke Quantum Unique Ergodicity} 
The main silent question of this paper is whether the system (\ref{Qsystem}) is quantum ergodic. Before discussing
this question, one is obliged to define a notion of quantum ergodicity. As a first approximation just copy the
classical definition, but replace each classical notion by its quantum counterpart. Namely, for every  $f \in \Ad$
and almost every quantum state $v \in S(\Hh)$ the following holds:

\begin{equation}\label{quantumergodicity}
 \lim_{N \rightarrow \infty} \frac{1}{N}\sum_{k=1}^{N}
  <v|\Pih(f^{A^k})v>= \int_{\T}f \ome.
\end{equation}

Unfortunately (\ref{quantumergodicity}) is literally not true. The
limit is never exactly equal the integral for a fixed $\hbar$.
Next we give a true statement which is a slight modification of
(\ref{quantumergodicity}), and is called {\it Hecke Quantum Unique
Ergodicity}. First rewrite (\ref{quantumergodicity}) in an
equivalent form. Using the Egorov identity (\ref{eqvquant3}) we have:

\begin{equation}\label{rw1}
  <v|\Pih(f^{A^k})v> = <v|\rhoh(A^k) \Pih(f) \rhoh(A^k)^{-1} v>
\end{equation}
\\
The elements $A^k$ runs inside the finite group $\GG$. Denote by
$\GrA \subseteq \GG$ the cyclic subgroup generated by $A$. It is
easy to see, using (\ref{rw1}), that:

\begin{equation}\label{rw2}
 \lim_{N \rightarrow \infty} \frac{1}{N}\sum_{k=1}^{N} <v|\Pih(f^{A^k})v>= \frac{1}{|\GrA|}\sum_{B \in \GrA} <v|\rhoh(B)
 \Pih(f) \rhoh(B)^{-1} v>
\end{equation}

Altogether (\ref{quantumergodicity}) can be written in the form:

\begin{equation}\label{rw3}
  \Av_{_{\GrA}} (<v|\Pih(f)v>) = \int_{\T}f \ome
\end{equation}
where $\Av_{_{\GrA}}$ denote the average  with respect to the group $\GrA$.
\\

\subsection{Hecke theory} Denote by $\CA = \{B \in \GG : BA=AB\}$ the centralizer of $A$ in
$\GG$. The finite group $\CA$ is an algebraic group. More
particulary, as an algebraic group, it is a torus. We call $\CA$
the {\it Hecke torus}. One has, $\GrA \subseteq \CA \subseteq
\GG$. Now, in (\ref{rw3}), average with respect to the group $\CA$
instead of the group $\GrA$. We assume the characteristic polynomial
$\mathrm{P}_A$ of $A$ is irreducible over $\Q$. Then the precise statement of the {\bf
Rudnick-Kurlberg conjecture} (which naturally generalize \cite{GH3}) is given in the following theorem:

\begin{theorem}[Hecke Quantum Unique Ergodicity]\label{GH3} Let $\hbar =
\frac{1}{p}, \r p>2$. For every $f \in \Ad$ and $v \in S(\Hh)$,
the following holds:
\begin{equation}\label{qheckerg}
  \left \| \Av_{_{\CA}} (<v|\Pih(f) v>) - \int_{\T}f \ome \right \| \leq \frac{C_{f}}{p^{n/2}},
\end{equation}

where $C_{f}$ is an explicit constant depends only on $f$.
\end{theorem}

The rest of this paper is devoted to proving Theorem \ref{GH3}.

\section{Proof of the Hecke Quantum Unique Ergodicity Conjecture}

The proof is given in two stages. The first stage is a preparation
stage and consists of mainly linear algebra considerations. We
massage statement (\ref{qheckerg}) in several steps into an
equivalent statement which will be better suited to our needs. In
the second stage we introduce the main part of the proof. Here we
invoke tools from algebraic geometry in the framework of
$\ell$-adic sheaves and $\ell$-adic cohomology (cf. \cite{M}).

\subsection{Preparation stage}

{\bf Step 1}. It is enough to prove (\ref{GH3}) for  $f$ a
character, $0 \neq f = \xi \in \Lm$. Because $\int_{\T}\xi \ome = 0$,
statement (\ref{qheckerg}) becomes:

\begin{equation}\label{qheckerg2}
  \left \| \Av_{_{\CA}}(<v|\Pih(\xi)v>) \right \|\leq \frac{C_\xi}{p^{n/2}}
\end{equation}

The statement for general $f \in \Ad$ follows
directly from the triangle inequality and the rapid decrease of the Fourier coefficients of $f$.\\

{\bf Step 2.} It is enough to prove (\ref{qheckerg2}) in case
$v \in S(\Hh)$ is a {\it Hecke} eigenvector. To be more  precise: the
{\it Hecke} torus $\CA$ acts semisimply on $\Hh$ via the
representation $\rhoh$, thus  $\Hh$ decomposes into  a direct sum of
character spaces:

\begin{equation}\label{decom}
  \Hh = \bigoplus_{\chi:\CA \lto \C^{\times}}\H_{\chi}.
\end{equation}

The sum in (\ref{decom}) is over multiplicative characters of the
torus $\CA$. For $v \in \H_{\chi}$, $B \in \CA$,

\begin{equation}\label{eigen}
\rhoh(B)v = \chi(B)v
\end{equation}

Taking $v \in \H_{\chi}$, statement (\ref{qheckerg2}) becomes:

\begin{equation}\label{qheckerg3}
  \|<v|\Pih(\xi)v> \| \leq \frac{C_\xi}{p^{n/2}}
\end{equation}

Here ${C_\xi} = 2^n$.\\

The averaged operator:

\begin{equation}
\frac{1}{|\CA|}\sum_{B \in \CA} \rhoh(B) \Pih(\xi) \rhoh(B)^{-1}
\end{equation}
is essentially diagonal in the Hecke base. Knowing this 
then statement (\ref{qheckerg2}) follows from (\ref{qheckerg3})
by invoking the triangle inequality.\\

{\bf Step 3.} Let $P_{\chi}:\Hh \lto \Hh$ be the orthogonal
projector on the eigenspace $\H_{\chi}$.

\begin{lemma}\label{lemma1}
For $\chi \neq 1$ dim $\H_{\chi}$ = 1.
\end{lemma}

Using lemma (\ref{lemma1}) we can rewrite (\ref{qheckerg3}) in the
form:

\begin{equation}\label{qheckerg4}
  \|\Tr(\Pih(\xi) P_{\chi}) \| \leq \frac{C_\xi}{p^{n/2}}.
\end{equation}

The projector $P_{\chi}$ can be defined in terms of the
representation $\rhoh$:

\begin{equation}\label{projector}
  P_{\chi} = \frac{1}{|\CA|}\sum_{B \in \CA} \chi(B) \rhoh(B).
\end{equation}

Now rewrite (\ref{qheckerg3}):

\begin{equation}\label{qheckerg4}
  \frac{1}{|\CA|} \left \| \sum_{B \in \CA} \Tr(\Pih(\xi) \rhoh(B))  \chi(B) \right \| \leq \frac{C_\xi}{p^{n/2}}
\end{equation}

Or noting that $|\CA| = p^n$, multiplying both sides of
(\ref{qheckerg4}) by $|\CA|$ we get:\\

\begin{theorem}[Hecke Quantum Unique Ergodicity (Restated)]\label{GH4}

Let $\hbar = \frac{1}{p}, \r p>2$ and let $\chi$ be a character of $\CA$. For every $0 \neq  \xi \in \Lm$ the following holds:

\begin{equation}\label{qheckerg5}
   \left \| \sum_{B \in \CA} \Tr(\Pih(\xi) \rhoh(B)) \chi(B) \right \| \leq C p^{n/2},
\end{equation}
where C = $2^n$.
\end{theorem}

We prove the Hecke ergodicity theorem in the form of Theorem
(\ref{GH4}).

\subsection {Proof of Theorem \ref{GH4}}

We prove Theorem \ref{GH4} using sheaf theoretic techniques.
Before diving into geometric considerations, we investigate further
the ingredients appearing in Theorem \ref{GH4}. For what
follows we fix a character  $\CHI : \CA \lto \C^{\times}$.\\

Denote by $F(\xi,B) = \Tr(\Pih(\xi) \rhoh(B))$, which is a function
of two variables, $F : \Lm \times \GG \lto \C$.\\

Denote by $\Lmm := \Lm / p \Lm$. The quotient lattice $\Lmm \simeq \mathbb{F}_{p}^{2n}$.
Set $\Y = \Lm \times \GG$,
and $\YY = \Lmm \times \GG$. One has the quotient map:

\begin{equation}\label{quotientmap}
  \Y \lto \YY.
\end{equation}

\begin{lemma}\label{factorization}
The function $F : \Y \lto \C$ factories through the quotient
$\YY$.
\[
\qtriangle[\Y`\YY`\C;`F`\overline{F}]
\]
\end{lemma}

Denote $\overline{F}$ also by $F$, and from now on $\YY$
will be considered as the default domain.\\

The function $F:\YY \lto \C$ is invariant under a certain group
action of $\GG$.  To be more precise, let $S \in \GG$,

\begin{equation}\label{invar1}
  \Tr( \Pih(\xi) \rhoh(B)) = \Tr( \rhoh(S) \Pih(\xi)
  \rhoh(S)^{-1} \rhoh(S) \rhoh(B) \rhoh(S)^{-1})
\end{equation}

Applying the Egorov identity (\ref{eqvquant3b}) and using the fact
that $\rhoh$ is a representation we get:

\begin{equation}\label{invar2}
  \Tr( \rhoh(S) \Pih(\xi) \rhoh(S)^{-1} \rhoh(S) \rhoh(B)
  \rhoh(S)^{-1})= \Tr(\Pih(S \xi) \rhoh(S B S^{-1}))
\end{equation}

Altogether we have:

\begin{equation}\label{invar3}
  F(\xi,B) = F(S \xi, SBS^{-1})
\end{equation}

Putting (\ref{invar3}) in a more diagrammatic form: there is an
action of $\GG$ on $\YY$ given by the following formula:

\begin{equation}\label{actionset}
  \begin{CD}
   \GG \times \YY    @>\alpha>> \YY \\
   (S,(\xi,B))       @>>>       (S \xi, SBS^{-1})
   \end{CD}
\end{equation}

Consider the following diagram:

\begin{equation}\label{actionset2}
  \begin{CD}
   \YY    @<pr<<    \GG \times \YY    @>\alpha>> \YY \\
  \end{CD}
\end{equation}

Where $pr$ is the projection on the $\YY$ variable. Formula
(\ref{invar3}) can be stated equivalently as:

\begin{equation}\label{invar4}
  \alpha^{*}(F) = pr^{*}(F),
\end{equation}

where $\alpha^{*}(F)$ and $pr^{*}(F)$  are the pullbacks of the function $F$ on $\YY$
via the maps $\alpha$ and $pr$  respectively. 

\subsection{Geometrization (Sheafification)}
What we are going to do next is to replace statement (\ref{GH4})
by a geometric statement, by which it will be implied. Going into
the geometric setting we replace the set $\YY$ by algebraic
variety and the functions $F$,$\r \CHI$ by sheaf
theoretic objects, also of a geometric flavor.\\

{\bf Step 1.} Notice that the set $\YY$ is not an arbitrary finite
set, but is the set of rational points of an  algebraic variety $\YYV$
defined over $\Fp$. To be more precise: $\YYV \simeq \FFp^{2n} \times
\GGV$. The variety $\YYV$ is equipped with an endomorphism:

\begin{equation}\label{frob}
  \Fr:\YYV \lto \YYV
\end{equation}

The endomorphism $\Fr$ is called Frobenius. The set $\YY$ is identified with the set of fixed points of Frobenius:

\begin{equation}\label{setfixed}
  \YY = \YYV^{\Fr} = \{ y \in \YYV : \Fr(y) = y \}.
\end{equation}

Note that the finite group $\GG$ is the set of rational points of
the  algebraic group $\GGV$. The finite quotient lattice $\Lmm$ is
the set of rational points of the affine space $\FFp^{2n}$.\\

Having all finite sets replaced by corresponding algebraic
varieties, we want to replace functions by sheaf theoretic
objects. This we do next.\\

{\bf step 2.} The following theorem, tells us about the appropriate sheaf theoretic object standing in place of
the function $F : \YY \lto \C$. Denote by $\Db (\YYV)$  the bounded derived category of {\it constructible}, $\ell$-adic
{\it Weil} sheaves (cf. \cite{M}).

\begin{theorem}[Deligne \cite{D1}]\label{deligne}

There exists a unique, up to an isomorphism, object $\SF \in
\Db(\YYV)$, which is of weight $w(\SF) \leq 0$ and is associated,
by {\bf sheaf-to-function correspondence}, to the function $F : \YY
\lto \C$:

\begin{equation}\label{sheaffunc}
  f^{\SF} = F
\end{equation}

\end{theorem}

We give here an intuitive explanation of Theorem {\ref{deligne}, part by part, as it was stated. By means of an
object $\SF \in \Db(\YYV)$ think of $\SF$ as a vector bundle over $\YYV$:

\begin{equation}\label{vectorbun}
  \begin{CD}
  \SF \\
  @VVV \\
  \YYV
\end{CD}
\end{equation}

The letter w in the notation $\Db$, means that $\SF$ is a {\it Weil} sheaf, namely it is a sheaf equipped with a lifting of the Frobenius:

\begin{equation}\label{lifting}
  \begin{CD}
   \SF  @>\Fr>>  \SF \\
   @VVV        @VVV \\
   \YYV @>\Fr>>  \YYV
  \end{CD}
\end{equation}

To be even more precise, think of $\SF$ not as a single vector
bundle, but as a complex of vector bundles over $\YYV$, $\SF =
\SF^{\bullet}$:

\begin{equation}\label{complex}
  \begin{CD}
   ...  @>d>> \SF^{-1}  @>d>> \SF^{0}  @>d>> \SF^{1}  @>d>> ...
   \end{CD}
\end{equation}
\\

$\SF^{\bullet}$ is equipped with a lifting of Frobenius:

\begin{equation}\label{liftingcomplex}
  \begin{CD}
   ...  @>d>> \SF^{-1}  @>d>> \SF^{0}  @>d>> \SF^{1}  @>d>> ... \\
    &    &      @V\Fr VV         @V\Fr VV           @V\Fr VV            \\
   ...  @>d>> \SF^{-1}  @>d>> \SF^{0}  @>d>> \SF^{1}  @>d>> ...
   \end{CD}
\end{equation}

Here the Frobenius commutes with the differentials.\\

Next we explain what is the meaning of the statement $w(\SF) \leq
0$. Let $y \in \YYV^{\Fr} = \YY$ be a fixed point of Frobenius.
Denote by $\SF_{y}$ the fiber of $\SF$ at the point $y$. Thinking
of $\SF$ as a complex of vector bundles, it is clear what one
means by taking the fiber at a point. The fiber $\SF_{y}$ is just a complex
of vector spaces. Because the point $y$ is fixed by the
Frobenius, it induces an endomorphism of $\SF_{y}$:

\begin{equation}\label{liftingcompy}
  \begin{CD}
   ...  @>d>> \SF^{-1}_{y}  @>d>> \SF^{0}_{y}  @>d>> \SF^{1}_{y}  @>d>> ... \\
    &    &      @V\Fr VV         @V\Fr VV           @V\Fr VV            \\
   ...  @>d>> \SF^{-1}_{y}  @>d>> \SF^{0}_{y}  @>d>> \SF^{1}_{y}  @>d>> ...
   \end{CD}
\end{equation}

The Frobenius acting as in (\ref{liftingcompy}) commutes with the differentials, so it induces an action on
cohomologies. For every $i \in \Z $ we have an endomorphism:

\begin{equation}\label{actcohom}
  \Fr:H^{i}(\SF_{y}) \lto H^{i}(\SF_{y})
\end{equation}

Saying an object $\SF$ has weight $w(\SF) \leq \omega$ means that for every point $y \in \YYV^{\Fr}$, and for every $i \in
\Z$ the absolute values of the eigenvalues of Frobenius acting on the $i$'th cohomology (\ref{actcohom}) satisfy:

\begin{equation}\label{weight}
 \left |\ev(\Fr \big{|}_{H^{i}(\SF_{y})}) \right | \leq \sqrt{p}^{\omega+i}
\end{equation}

In our case $\omega = 0$ and so:

\begin{equation}\label{weight}
 \left |\ev(\Fr \big{|}_{H^{i}(\SF_{y})}) \right | \leq \sqrt{p}^{i}
\end{equation}

The last part of Theorem \ref{deligne} concerns a function
$f^{\SF}: \YY \lto \C$ associated to the sheaf $\SF$. To define
$f^{\SF}$, we have to describe its value at every point $y
\in \YY$. Let $y \in \YY = \YYV^{\Fr}$. Frobenius acts on the
cohomologies of the fiber $\SF_{y}$  (cf. (\ref{actcohom})). Now put:

\begin{equation}\label{eulerchar}
  f^{\SF}(y) = \sum_{i \in \Z} (-1)^i \Tr(\Fr \big{|}_{H^i(\SF_y)}).
\end{equation}

In words: $f^{\SF}(y)$ is the alternating sum of traces of Frobenius acting on the cohomologies of the fiber
$\SF_y$. We call this sum the {\it Euler characteristic} of Frobenius and denote it by:

\begin{equation}\label{eulerchar2}
  f^{\SF}(y) = \chi_{_{\Fr}}(\SF_y)
\end{equation}

Theorem \ref{deligne} asserts that $f^{\SF}$ is the function $F$ defined earlier. Associating the function
$f^{\SF}$ on the set $\YYV^{\Fr}$ to the sheaf $\SF$ on $\YYV$ is a particular case  of a general procedure called {\it Grothendieck's
Sheaf-to-Function Correspondence} \cite{G}. Because we are going to use this procedure later,
next we spend some time explaining it in greater generality (see also \cite{Ga}).\\

\underline{\bf Grothendieck's Sheaf-to-Function Correspondence}.\\

Let $\overline{X}$ be an algebraic variety defined over $\Fp$. This means that there exists a Frobenius
endomorphism:

\begin{equation}\label{frob}
  \Fr : \overline{X} \lto \overline{X}
\end{equation}

The set $ X = \overline{X}^{\Fr}$ is called the set of rational points of $\overline{X}$. Let ${\cal L} \in
\Db(\overline{X})$ be a Weil sheaf. One can associate to ${\cal L}$ a function $f^{\cal L}$ on the set $X$ by the
following formula:

\begin{equation}\label{stfc}
  f^{\cal L}(x) = \sum_{i \in \Z} (-1)^i \Tr(\Fr \big{|}_{H^i({\cal L}_x)})
\end{equation}

This procedure is called {\it Sheaf-to-Function correspondence}. Next we describe some important functorial
properties of the procedure:\\

 Let $\overline{X}_{1}$, $\overline{X}_{2}$ be algebraic varieties defined over $\Fp$. Let $X_{1} =
\overline{X}^{\Fr}_{1}$, and $X_{2} = \overline{X}^{\Fr}_{2}$ be the corresponding sets of rational points. Let $\pi
: \overline{X}_{1} \lto \overline{X}_{2}$ be a morphism of algebraic varieties. Denote also by
$ \pi : X_{1} \lto X_{2}$ the induced map on the level of sets.\\

{\bf First statement}. Say we have a sheaf ${\cal L} \in
\Db(\overline{X}_{2})$. The following holds:

\begin{equation}\label{sfc1}
 f^{\pi^{*}({\cal L})} = \pi^{*}(f^{\cal L}),
\end{equation}

where on the function level $\pi^{*}$ is just pull back of functions. On the sheaf theoretic level $\pi^{*}$ is the
pull-back functor of sheaves (think of pulling back a vector bundle). Equation (\ref{sfc1}) says that  
{\it Sheaf-to-Function Correspondence} commutes with the operation of pull back.\\

{\bf Second statement}. Say we have a sheaf ${\cal L} \in
\Db(\overline{X}_{1})$.  The following holds:

\begin{equation}\label{sfc2}
 f^{\pi_{!}({\cal L})} = \pi_{!}(f^{\cal L}),
\end{equation}

where on the function level $\pi_{!}$ means to sum up the values of the function along the fibers of the map
$\pi$. On the sheaf theoretic level $\pi_{!}$ is compact integration of sheaves ( here we have no analogue under
the vector bundle interpretation). Equation (\ref{sfc2}) says that  
{\it Sheaf-to-Function Correspondence} commutes with integration.\\

{ \bf Third statement}. Say we have two sheaves ${\cal L}_{1},
{\cal L}_{2} \in \Db(\overline{X}_{1})$. The following holds:

\begin{equation}\label{sfc3}
 f^{{\cal L}_{1} \otimes {\cal L}_{2}} = f^{{\cal L}_{1}} \cdot f^{{\cal
 L}_{2}}
\end{equation}

In words: {\it Sheaf-to-Function Correspondence} takes tensor product of sheaves to multiplication of the
corresponding functions.

\subsection{Geometric statement}

Fix an element $0 \neq \XI \in \Lm$. Denote By $\i_XI$ the inclusion map:

\begin{equation}\label{map1}
  \i_XI: \{\XI\} \times \CA \lto \YY
\end{equation}

and the canonical projection:

\begin{equation}\label{map2}
  \p_XI: \{\XI\} \times \CA \lto pt
\end{equation}

Going back to Theorem \ref{GH4}, and putting its content in a
diagrammatic form, we obtain the following inequality:\\

\begin{equation}\label{Diagramatic1}
  \| {\p_XI}_{!}( \i_XI^{*}(F) \cdot \CHI) \| \leq 2^n p^{n/2}
\end{equation}

In words: taking the function $F : \YY \lto \C$.
\begin{itemize}
  \item Restrict $F$ to $\{\XI\} \times \CA$ and get $\i_XI^{*}(F)$
  \item Multiply $\i_XI^{*}(F)$ by the character $\CHI$ to get
  $\i_XI^{*}(F) \cdot \CHI$
  \item Integrate $\i_XI^{*}(F)\CHI$ to the point, that is sum all the
  values, to get a scalar $a_{\CHI} = {\p_XI}_{!}(\i_XI^{*}(F) \cdot \CHI)$
\end{itemize}
Theorem \ref{GH4} claims that the scalar $a_{\CHI}$ is of absolute value less than $2^n p^{n/2}$.\\

Now do the same thing on the geometric level. We have the closed embedding:

\begin{equation}\label{map1b}
  \i_XI: \{\XI\} \times \CAV \lto \YYV
\end{equation}

And the canonical projection:

\begin{equation}\label{map2b}
  \p_XI: \{\XI\} \times \CAV \lto pt
\end{equation}

Take the sheaf $\SF$ on $\YYV$,

\begin{itemize}
  \item Pull-back $\SF$ to the closed subvariety $\{ \XI \} \times
  \CAV$, to get the sheaf $\i_XI^{*}(\SF)$.
  \item Take a tensor product of $\i_XI^{*}(\SF)$ with the Kummer
  sheaf ${\SL}_{\CHI}$, to get $\i_XI^{*}(\SF) \otimes {\SL}_{\CHI}$
  \item Integrate $\i_XI^{*}(\SF) \otimes {\SL}_{\CHI}$ to the point, to get the sheaf ${\p_XI}_{!}(\i_XI^{*}(\SF) \otimes {\SL}_{\CHI})$ on the point.
\end{itemize}

The Kummer sheaf ${\SL}_{\CHI}$ is associated via {\it
Sheaf-to-Function Correspondence} to the character $\CHI$.\\

Recall that {\it Sheaf-to-Function Correspondence} commutes both with pullback (\ref{sfc1}), with integration
(\ref{sfc2}), and takes tensor product of sheaves to multiplication of functions (\ref {sfc3}). This means that it
commutes the operations done on the level of sheaves with those done on the level of functions. The following
diagram describe pictorially, what has been said so far:

\begin{equation}\label{diag1}
\begin{CD}
   \SF                                        @>\chi_{_\Fr}>>              & F               \\
   @A\i_XIAA            &                                            @A \i_XI AA            \\
  \i_XI^{*}(\SF) \otimes {\SL}_{\CHI}      @>\chi_{_\Fr}>>    & \i_XI^{*}(F) \cdot \CHI    \\
   @V \p_XI VV           &                                             @V \p_XI VV        \\
 {\p_XI}_{!}(\i_XI^{*}(\SF) \otimes {\SL}_{\CHI})  @>\chi_{_\Fr}>>     &  {\p_XI}_{!}(\i_XI^{*}(F)\cdot \CHI)
\end{CD}
\end{equation}

Denote by  $\SG = {\p_XI}_{!}(i^*(\SF) \otimes \SL_{\CHI})$. It is an object in $\Db(pt)$. This means it is merely
a complex of vector spaces, $\SG = \SG^{\bullet}$, together with an action of Frobenius:

\begin{equation}\label{gcomplex}
  \begin{CD}
   ...  @>d>> \SG^{-1}  @>d>> \SG^{0}  @>d>> \SG^{1}  @>d>> ... \\
    &    &      @V\Fr VV         @V\Fr VV           @V\Fr VV            \\
   ...  @>d>> \SG^{-1}  @>d>> \SG^{0}  @>d>> \SG^{1}  @>d>> ...
   \end{CD}
\end{equation}

The complex $\SG^{\bullet}$ is associated by {\it Sheaf-to-Function correspondence} to the scalar $a_{\CHI}$: 

\begin{equation}\label{geulerchar}
  a_{\CHI} = \sum_{i \in \Z} (-1)^i \Tr(\Fr \big{|}_{H^i(\SG)})
\end{equation}

At last we can give the geometric statement  about $\SG$,  which will imply Theorem \ref{GH4}.

\begin{lemma}[Geometric Lemma]\label{vanishing}
Let $\SG = {\p_XI}_{!}(\i_XI^*(\SF) \otimes \SL_{\CHI}))$. There exist a natural number $0\leq \mchi \leq n$ such that:
\begin{enumerate}

\item All cohomologies $H^{i}(\SG)$ vanish except for $i=n+\mchi$.
Moreover $H^i(\SG)$ is a $2^n$-dimensional vector space.

\item The sheaf $\SG$ has weight $w(\SG) \leq -2\mchi$.

\end{enumerate}
\end{lemma}

Theorem \ref{GH4} now follows easily. Having $w(\SG) \leq -2\mchi$
means that:

\begin{equation}\label{weight2}
 \left |\ev(\Fr \big{|}_{H^{i}(\SG)}) \right | \leq \sqrt{p}^{i - 2\mchi}.
\end{equation}

By Lemma \ref{vanishing} only the cohomology $H^{n+\mchi}(\SG)$ does not vanish and it is $2^n$-dimensional. The eigenvalues
of Frobenius acting on $H^{n+ \mchi}(\SG)$ are of absolute value $ \leq \sqrt{p}^{n-\mchi}$, (\ref{weight2}). It now follows, using
formula (\ref{geulerchar}), that:

\begin{equation}\label{final}
  | a_{\CHI} | \leq 2^n p^{n/2}
\end{equation}

for all $\chi$.\\

The rest of the paper is devoted to the proof of Lemma \ref{vanishing}.

\subsection{Proof of the Geometric Lemma}

The proof will be given in several steps, reducing the Geometric Lemma to the case n=1.\\

{\bf Step 1.} The sheaf $\SF$ is $\GGV$-equivariant sheaf. 

Recall that the function $F:\YY \lto \C$ is invariant
under a group action of $\GG$ on $\YY$. One can define the
analogue group action in the geometric setting. Here the algebraic
group $\GGV$ acts on the variety $\YYV$. The formulas are the same
as those given in (\ref{actionset}):

\begin{equation}\label{actionsetb}
  \begin{CD}
   \GGV \times \YYV    @>\alpha>> \YYV \\
   (S,(\xi,B))       @>>>       (S \xi, SBS^{-1})
   \end{CD}
\end{equation}

It turns out that the invariance property of the function $F$ is a
manifestation of a finer geometric phenomena. Namely the sheaf
$\SF$ is equivariant with respect to the action $\alpha$. More
precisely, we have the diagram:

\begin{equation}\label{actionset2b}
  \begin{CD}
   \YYV    @<pr<<    \GGV \times \YYV    @>\alpha>> \YYV \\
  \end{CD}
\end{equation}

There exists an isomorphism $\theta$:

\begin{equation}\label{invar4b}
   \theta : \alpha^{*}(\SF) \simeq pr^{*}(\SF)
\end{equation}
\\

{\bf Step 2}. All tori in $\GGV$ are conjugated over $\Fq$, where $q=2n$. In particular there exists an element
$\S0 \in Sp(2n,\Fq)$ conjugating the {\it Hecke} torus $\CAV \subset \GGV$ with the standard torus $\TTV \subset \GGV$,

\begin{equation}\label{conj}
\S0 \CAV \S0^{-1} = \TTV \
\end{equation}

The standard torus is:

\begin{equation}\label{stdtorus}
\TTV = \left \{
\begin{pmatrix}
a_1 &          &        &        &     &  \\
    & a_1^{-1} &        &        &     &  \\
    &          & \ddots &        &     &  \\
    &          &        & \ddots &     &  \\
    &          &        &        & a_n &  \\
    &          &        &        &     & a_n^{-1}\\
\end{pmatrix} : a_i \in \FFp^{\times} 
\right \}
\end{equation}

The situation is displayed in the following diagram:

\begin{equation}\label{diag2}
  \begin{CD}
    \LmmV \times \GGV      @>\alpha_{_{\S0}}>>        \LmmV \times \GGV   \\
      @A\i_XIAA                                   @A \iXII AA \\
    \{ \XI \} \times \CAV    @>\alpha_{_{\S0}}>>      \{ \XII \} \times \TTV \\
     @V \p_XI VV                                        @V \pXII VV                \\
     pt                     @=                        pt
 \end{CD}
\end{equation}

Where $\XII = \S0 \cdot \XI$, and $\alpha_{_{\S0}}$ is the restriction of the action $\alpha$, (\ref{actionsetb})
to the element $\S0$.\\

{\bf Step 3.} Denote by $\SG'$ the object in $\Db(pt)$ defined by:
\begin{equation}
  \SG':= {\pXII}_{!}(\iXII^*(\SF) \otimes {\alpha_{_{\S0}}}_{_{!}}(\SL_{\CHI}))
\end{equation}

Then $\SG$ and $\SG'$ are isomorphic as  {\it ``quasi Weil sheaves''}. Namely there exist an isomorphism:
\begin{equation}\label{qWiso}
\SG \iso \SG'
\end{equation}
which commutes with the action of $\Fr _q := \Fr^{2n}$.\\

By base change:

\begin{equation}\label{seq1}
  {\p_XI}_{!}(\i_XI^*(\SF) \otimes \SL_{\CHI}) \iso {\pXII}_{!} ({\alpha_{_{\S0}}}_{_{!}}(\i_XI^*(\SF) \otimes \SL_{\CHI}))
\end{equation}

By usual property:

\begin{equation}\label{seq2}
 {\pXII}_{!} ({\alpha_{_{\S0}}}_{_{!}}(\i_XI^*(\SF) \otimes \SL_{\CHI})) \iso {\pXII}_{!} ({\alpha_{_{\S0}}}_{_{!}}(\i_XI^*(\SF)) \otimes
{\alpha_{_{\S0}}}_{_{!}}(\SL_{\CHI})))
\end{equation}

By base change:

\begin{equation}\label{seq3}
 {\alpha_{_{\S0}}}_{_{!}}(\i_XI^*(\SF)) \iso \iXII^*({\alpha_{_{\S0}}}_{_{!}}(\SF))
\end{equation}

By the equivariance property of the sheaf $\SF$, (\ref{invar4b}), we have an isomorphism:

\begin{equation}\label{seq4}
   \theta_{_{\S0}} : {\alpha_{_{\S0}}}_{_{!}}(\SF) \simeq \SF
\end{equation}

Combining (\ref{seq1}), (\ref{seq2}), (\ref{seq3}), (\ref{seq4})  we get:

\begin{equation}\label{seq5}
   {\p_XI}_{!}(\i_XI^*(\SF) \otimes \SL_{\CHI}) \iso  {\pXII}_{!}(\iXII^*(\SF) \otimes {\alpha_{_{\S0}}}_{_{!}}(\SL_{\CHI}))
\end{equation}

as claimed.\\

{\bf Step 4.} It is enough to prove the Geometric Lemma for the sheaf $\SG'$.
Knowing the weight $w(\SG')$, then using the isomorphism (\ref{qWiso}) we know the weight $w(\SG)$ with respect to the action 
of $\Fr_q$ and hence also for the action of $\Fr$.\\

{\bf Step 5. (Factorization).} The sheaf $\SG'$ factories:

\begin{equation}\label{fac}
\SG'\iso \otimes_{j=1}^n \SG'_j
\end{equation}

where the sheaves  $\SG'_j$ are the sheaves on $pt$ corresponds to the case n=1 over $\Fq$.
This fact follows from factorization properties of the sheaves $\SF$ and $\SL_{\CHI}$, which we are able to show over
the torus $\TTV$ where we have explicit formulas for the sheaves involved. \\

{\bf Step 6.} The geometric Lemma holds for any of the sheaves $\SG'_j$. 

First we compute the sheaf $\iXII^*(\SF)$.\\

Denote by $\psi:\Fq \to \C^{\times}$ the additive character $\psi(t) = e^\frac{2 \pi i \cdot tr(t)}{p}$. 
Write $\XII$ in the coordinates of the standard
basis of $\Fq^2$, say $ \XII = (\lam,\mu)$. Let $\SF_{\psi}$ be the Artin-Schreier sheaf corresponding to the
character $\psi$. Consider the morphism:

\begin{equation}\label{map1}
f: \TTV - \{ Id \} \lto \FFp
\end{equation}

given by the formula:

\begin{equation}\label{map2}
  f  \left ( \begin{pmatrix} a & 0 \\ 0 & a^{-1} \end{pmatrix}\right )  = \half(\lam \cdot \mu) \cdot \frac{1+a}{1-a}
\end{equation}

We have:

\begin{equation}\label{map3}
\iXII^*(\SF) = f^*(\SF_{\psi})
\end{equation}

Note that the sheaf $f^*(\SF_{\psi})$ although apriori defined over $\TTV - \{ Id \}$, has all its
extensions to $\TTV$ coincide. So in (\ref{map3}) we consider some extension to $\TTV$, and there is no ambiguity
here.\\

About the sheaf ${\alpha_{_{\S0}}}_{_{!}}(\SL_{\CHI})$, it is
isomorphic to an explicit Kummer sheaf on $\TTV$ correspond to a character  $\CHI':\TTV(\Fq) \lto \C^{\times}$.
Altogether we get that $\SG'$ is a kind of a Gauss-sum sheaf. More precisely this is a sheaf given by explicit formula.
By direct computation we prove the Geometric Lemma (\ref{vanishing}) for this sheaf. We found that $\mchi = 1$ if
$\CHI'$ is trivial and $\mchi = 0$ otherwise.\\

Now using the property:
\[
w(\SG_1 \otimes \SG_2) = w(\SG_1) + w(\SG_2)
\]

and Kunneth formula:

\[
H^i(\SG_1 \otimes \SG_2) = \sum\limits_{k+l = i} H^k(\SG_1) \otimes H^l(\SG_2)
\]

The proof is completed. $\EProof$

\bigskip \noindent
S.G,\\
Tel Aviv University, Israel. \\
{\it E-mail: shamgar@math.tau.ac.il}

\bigskip \noindent
R.H,\\
Tel Aviv University, Israel. \\
{\it E-mail: hadani@post.tau.ac.il}

\bigskip

\center{\it  September 1 2004.}

\end{document}

%% file: diagram.tex
\makeatletter
 
\def\diagram{\m@th\leftwidth=\z@ \rightwidth=\z@ \topheight=\z@
\botheight=\z@ \setbox\@picbox\hbox\bgroup}
 
\def\enddiagram{\egroup\wd\@picbox\rightwidth\unitlength
\ht\@picbox\topheight\unitlength \dp\@picbox\botheight\unitlength
\hskip\leftwidth\unitlength\box\@picbox}
 
\def\bfig{\begin{diagram}}
\def\efig{\end{diagram}}
\newcount\wideness \newcount\leftwidth \newcount\rightwidth
\newcount\highness \newcount\topheight \newcount\botheight
 
\def\ratchet#1#2{\ifnum#1<#2 \global #1=#2 \fi}
 
\def\putbox(#1,#2)#3{%
\horsize{\wideness}{#3} \divide\wideness by 2
{\advance\wideness by #1 \ratchet{\rightwidth}{\wideness}}
{\advance\wideness by -#1 \ratchet{\leftwidth}{\wideness}}
\vertsize{\highness}{#3} \divide\highness by 2
{\advance\highness by #2 \ratchet{\topheight}{\highness}}
{\advance\highness by -#2 \ratchet{\botheight}{\highness}}
\put(#1,#2){\makebox(0,0){$#3$}}}
 
\def\putlbox(#1,#2)#3{%
\horsize{\wideness}{#3}
{\advance\wideness by #1 \ratchet{\rightwidth}{\wideness}}
{\ratchet{\leftwidth}{-#1}}
\vertsize{\highness}{#3} \divide\highness by 2
{\advance\highness by #2 \ratchet{\topheight}{\highness}}
{\advance\highness by -#2 \ratchet{\botheight}{\highness}}
\put(#1,#2){\makebox(0,0)[l]{$#3$}}}
 
\def\putrbox(#1,#2)#3{%
\horsize{\wideness}{#3}
{\ratchet{\rightwidth}{#1}}
{\advance\wideness by -#1 \ratchet{\leftwidth}{\wideness}}
\vertsize{\highness}{#3} \divide\highness by 2
{\advance\highness by #2 \ratchet{\topheight}{\highness}}
{\advance\highness by -#2 \ratchet{\botheight}{\highness}}
\put(#1,#2){\makebox(0,0)[r]{$#3$}}}

\def\adjust[#1]{} 
 
\newcount \coefa
\newcount \coefb
\newcount \coefc
\newcount\tempcounta
\newcount\tempcountb
\newcount\tempcountc
\newcount\tempcountd
\newcount\xext
\newcount\yext
\newcount\xoff
\newcount\yoff
\newcount\gap%
\newcount\arrowtypea
\newcount\arrowtypeb
\newcount\arrowtypec
\newcount\arrowtyped
\newcount\arrowtypee
\newcount\height
\newcount\width
\newcount\xpos
\newcount\ypos
\newcount\run
\newcount\rise
\newcount\arrowlength
\newcount\halflength
\newcount\arrowtype
\newdimen\tempdimen
\newdimen\xlen
\newdimen\ylen
\newsavebox{\tempboxa}%
\newsavebox{\tempboxb}%
\newsavebox{\tempboxc}%
 
\newdimen\w@dth
 
\def\setw@dth#1#2{\setbox\z@\hbox{\m@th$#1$}\w@dth=\wd\z@
\setbox\@ne\hbox{\m@th$#2$}\ifnum\w@dth<\wd\@ne \w@dth=\wd\@ne \fi
\advance\w@dth by 1.2em}
 
\def\t@^#1_#2{\allowbreak\def\n@one{#1}\def\n@two{#2}\mathrel
{\setw@dth{#1}{#2}
\mathop{\hbox to \w@dth{\rightarrowfill}}\limits
\ifx\n@one\empty\else ^{\box\z@}\fi
\ifx\n@two\empty\else _{\box\@ne}\fi}}
\def\t@@^#1{\@ifnextchar_{\t@^{#1}}{\t@^{#1}_{}}}
\def\to{\@ifnextchar^{\t@@}{\t@@^{}}}
 
\def\t@left^#1_#2{\def\n@one{#1}\def\n@two{#2}\mathrel{\setw@dth{#1}{#2}
\mathop{\hbox to \w@dth{\leftarrowfill}}\limits
\ifx\n@one\empty\else ^{\box\z@}\fi
\ifx\n@two\empty\else _{\box\@ne}\fi}}
\def\t@@left^#1{\@ifnextchar_{\t@left^{#1}}{\t@left^{#1}_{}}}
\def\toleft{\@ifnextchar^{\t@@left}{\t@@left^{}}}
 
\def\two@^#1_#2{\allowbreak
\def\n@one{#1}\def\n@two{#2}\mathrel{\setw@dth{#1}{#2}
\mathop{\vcenter{\lineskip\z@\baselineskip\z@
                 \hbox to \w@dth{\rightarrowfill}%
                 \hbox to \w@dth{\rightarrowfill}}%
       }\limits
\ifx\n@one\empty\else ^{\box\z@}\fi
\ifx\n@two\empty\else _{\box\@ne}\fi}}
\def\tw@@^#1{\@ifnextchar _{\two@^{#1}}{\two@^{#1}_{}}}
\def\two{\@ifnextchar ^{\tw@@}{\tw@@^{}}}
 
\def\tofr@^#1_#2{\def\n@one{#1}\def\n@two{#2}\mathrel{\setw@dth{#1}{#2}
\mathop{\vcenter{\hbox to \w@dth{\rightarrowfill}\kern-1.7ex
                 \hbox to \w@dth{\leftarrowfill}}%
       }\limits
\ifx\n@one\empty\else ^{\box\z@}\fi
\ifx\n@two\empty\else _{\box\@ne}\fi}}
\def\t@fr@^#1{\@ifnextchar_ {\tofr@^{#1}}{\tofr@^{#1}_{}}}
\def\tofro{\@ifnextchar^ {\t@fr@}{\t@fr@^{}}}

\def\mon{\mathop{\m@th\hbox to
      14.6\P@{\lasyb\char'51\hskip-2.1\P@$\arrext$\hss
$\mathord\rightarrow$}}\limits} 
\def\leftmono{\mathrel{\m@th\hbox to
14.6\P@{$\mathord\leftarrow$\hss$\arrext$\hskip-2.1\P@\lasyb\char'50%
}}\limits} 
\mathchardef\arrext="0200       

\def\settypes(#1,#2,#3){\arrowtypea#1 \arrowtypeb#2 \arrowtypec#3}
\def\settoheight#1#2{\setbox\@tempboxa\hbox{#2}#1\ht\@tempboxa\relax}%
\def\settodepth#1#2{\setbox\@tempboxa\hbox{#2}#1\dp\@tempboxa\relax}%
\def\settokens`#1`#2`#3`#4`{%
     \def\tokena{#1}\def\tokenb{#2}\def\tokenc{#3}\def\tokend{#4}}
\def\setsqparms[#1`#2`#3`#4;#5`#6]{%
\arrowtypea #1
\arrowtypeb #2
\arrowtypec #3
\arrowtyped #4
\width #5
\height #6
}
\def\setpos(#1,#2){\xpos=#1 \ypos#2}

\def\settriparms[#1`#2`#3;#4]{\settripairparms[#1`#2`#3`1`1;#4]}%
 
\def\settripairparms[#1`#2`#3`#4`#5;#6]{%
\arrowtypea #1
\arrowtypeb #2
\arrowtypec #3
\arrowtyped #4
\arrowtypee #5
\width #6
\height #6
}
 
\def\resetparms{\settripairparms[1`1`1`1`1;500]\width 500}
 
\resetparms
 
\def\mvector(#1,#2)#3{
\put(0,0){\vector(#1,#2){#3}}%
\put(0,0){\vector(#1,#2){26}}%
}
\def\evector(#1,#2)#3{{
\arrowlength #3
\put(0,0){\vector(#1,#2){\arrowlength}}%
\advance \arrowlength by-30
\put(0,0){\vector(#1,#2){\arrowlength}}%
}}
 
\def\horsize#1#2{%
\settowidth{\tempdimen}{$#2$}%
#1=\tempdimen
\divide #1 by\unitlength
}
 
\def\vertsize#1#2{%
\settoheight{\tempdimen}{$#2$}%
#1=\tempdimen
\settodepth{\tempdimen}{$#2$}%
\advance #1 by\tempdimen
\divide #1 by\unitlength
}
 
\def\putvector(#1,#2)(#3,#4)#5#6{{%
\ifnum3<\arrowtype
\putdashvector(#1,#2)(#3,#4)#5\arrowtype
\else
\ifnum\arrowtype<-3
\putdashvector(#1,#2)(#3,#4)#5\arrowtype
\else
\xpos=#1
\ypos=#2
\run=#3
\rise=#4
\arrowlength=#5
\ifnum \arrowtype<0
    \ifnum \run=0
        \advance \ypos by-\arrowlength
    \else
        \tempcounta \arrowlength
        \multiply \tempcounta by\rise
        \divide \tempcounta by\run
        \ifnum\run>0
            \advance \xpos by\arrowlength
            \advance \ypos by\tempcounta
        \else
            \advance \xpos by-\arrowlength
            \advance \ypos by-\tempcounta
        \fi
    \fi
    \multiply \arrowtype by-1
    \multiply \rise by-1
    \multiply \run by-1
\fi
\ifcase \arrowtype
\or \put(\xpos,\ypos){\vector(\run,\rise){\arrowlength}}%
\or \put(\xpos,\ypos){\mvector(\run,\rise)\arrowlength}%
\or \put(\xpos,\ypos){\evector(\run,\rise){\arrowlength}}%
\fi\fi\fi
}}
 
\def\putsplitvector(#1,#2)#3#4{
\xpos #1
\ypos #2
\arrowtype #4
\halflength #3
\arrowlength #3
\gap 140
\advance \halflength by-\gap
\divide \halflength by2
\ifnum\arrowtype>0
   \ifcase \arrowtype
   \or \put(\xpos,\ypos){\line(0,-1){\halflength}}%
       \advance\ypos by-\halflength
       \advance\ypos by-\gap
       \put(\xpos,\ypos){\vector(0,-1){\halflength}}%
   \or \put(\xpos,\ypos){\line(0,-1)\halflength}%
       \put(\xpos,\ypos){\vector(0,-1)3}%
       \advance\ypos by-\halflength
       \advance\ypos by-\gap
       \put(\xpos,\ypos){\vector(0,-1){\halflength}}%
   \or \put(\xpos,\ypos){\line(0,-1)\halflength}%
       \advance\ypos by-\halflength
       \advance\ypos by-\gap
       \put(\xpos,\ypos){\evector(0,-1){\halflength}}%
   \fi
\else \arrowtype=-\arrowtype
   \ifcase\arrowtype
   \or \advance \ypos by-\arrowlength
       \put(\xpos,\ypos){\line(0,1){\halflength}}%
       \advance\ypos by\halflength
       \advance\ypos by\gap
       \put(\xpos,\ypos){\vector(0,1){\halflength}}%
   \or \advance \ypos by-\arrowlength
       \put(\xpos,\ypos){\line(0,1)\halflength}%
       \put(\xpos,\ypos){\vector(0,1)3}%
       \advance\ypos by\halflength
       \advance\ypos by\gap
       \put(\xpos,\ypos){\vector(0,1){\halflength}}%
   \or \advance \ypos by-\arrowlength
       \put(\xpos,\ypos){\line(0,1)\halflength}%
       \advance\ypos by\halflength
       \advance\ypos by\gap
       \put(\xpos,\ypos){\evector(0,1){\halflength}}%
   \fi
\fi
}
 
\def\putmorphism(#1)(#2,#3)[#4`#5`#6]#7#8#9{{%
\run #2
\rise #3
\ifnum\rise=0
  \puthmorphism(#1)[#4`#5`#6]{#7}{#8}#9%
\else\ifnum\run=0
  \putvmorphism(#1)[#4`#5`#6]{#7}{#8}#9%
\else
\setpos(#1)%
\arrowlength #7
\arrowtype #8
\ifnum\run=0
\else\ifnum\rise=0
\else
\ifnum\run>0
    \coefa=1
\else
   \coefa=-1
\fi
\ifnum\arrowtype>0
   \coefb=0
   \coefc=-1
\else
   \coefb=\coefa
   \coefc=1
   \arrowtype=-\arrowtype
\fi
\width=2
\multiply \width by\run
\divide \width by\rise
\ifnum \width<0  \width=-\width\fi
\advance\width by60
\if l#9 \width=-\width\fi
\putbox(\xpos,\ypos){#4}
{\multiply \coefa by\arrowlength
\advance\xpos by\coefa
\multiply \coefa by\rise
\divide \coefa by\run
\advance \ypos by\coefa
\putbox(\xpos,\ypos){#5} }%
{\multiply \coefa by\arrowlength
\divide \coefa by2
\advance \xpos by\coefa
\advance \xpos by\width
\multiply \coefa by\rise
\divide \coefa by\run
\advance \ypos by\coefa
\if l#9%
   \putrbox(\xpos,\ypos){#6}%
\else\if r#9%
   \putlbox(\xpos,\ypos){#6}%
\fi\fi }%
{\multiply \rise by-\coefc
\multiply \run by-\coefc
\multiply \coefb by\arrowlength
\advance \xpos by\coefb
\multiply \coefb by\rise
\divide \coefb by\run
\advance \ypos by\coefb
\multiply \coefc by70
\advance \ypos by\coefc
\multiply \coefc by\run
\divide \coefc by\rise
\advance \xpos by\coefc
\multiply \coefa by140
\multiply \coefa by\run
\divide \coefa by\rise
\advance \arrowlength by\coefa
\ifcase\arrowtype
\or \put(\xpos,\ypos){\vector(\run,\rise){\arrowlength}}%
\or \put(\xpos,\ypos){\mvector(\run,\rise){\arrowlength}}%
\or \put(\xpos,\ypos){\evector(\run,\rise){\arrowlength}}%
\fi}\fi\fi\fi\fi}}

\newcount\numbdashes \newcount\lengthdash \newcount\increment
 
\def\howmanydashes{
\numbdashes=\arrowlength \lengthdash=40
\divide\numbdashes by \lengthdash
\lengthdash=\arrowlength
\divide\lengthdash by \numbdashes
\increment=\lengthdash
\multiply\lengthdash by 3
\divide\lengthdash by 5
}
 
\def\putdashvector(#1)(#2,#3)#4#5{%
\ifnum#3=0 \putdashhvector(#1){#4}#5
\else
\ifnum#2=0
\putdashvvector(#1){#4}#5\fi\fi}
 
\def\putdashhvector(#1,#2)#3#4{{%
\arrowlength=#3 \howmanydashes
\multiput(#1,#2)(\increment,0){\numbdashes}%
{\vrule height .4pt width \lengthdash\unitlength}
\arrowtype=#4 \xpos=#1
\ifnum\arrowtype<0 \advance\arrowtype by 7 \fi
\ifcase\arrowtype
\or \advance\xpos by 10
    \put(\xpos,#2){\vector(-1,0){\lengthdash}}
    \advance\xpos by 40
    \put(\xpos,#2){\vector(-1,0){\lengthdash}}
\or \advance \xpos by 10
    \put(\xpos,#2){\vector(-1,0){\lengthdash}}
    \advance\xpos by  \arrowlength
    \advance\xpos by  -50
    \put(\xpos,#2){\vector(-1,0){\lengthdash}}
\or \advance\xpos by 10
    \put(\xpos,#2){\vector(-1,0){\lengthdash}}
\or \advance\xpos by \arrowlength
    \advance\xpos by -\lengthdash
    \put(\xpos,#2){\vector(1,0){\lengthdash}}
\or {\advance\xpos by 10
    \put(\xpos,#2){\vector(1,0){\lengthdash}}}
    \advance\xpos by \arrowlength
    \advance\xpos by -\lengthdash
    \put(\xpos,#2){\vector(1,0){\lengthdash}}
\or \advance\xpos by \arrowlength
    \advance\xpos by -\lengthdash
    \put(\xpos,#2){\vector(1,0){\lengthdash}}
    \advance\xpos by -40
    \put(\xpos,#2){\vector(1,0){\lengthdash}}
   \fi
}}
 
\def\putdashvvector(#1,#2)#3#4{{%
\arrowlength=#3 \howmanydashes
\ypos=#2 \advance\ypos by -\arrowlength
\multiput(#1,#2)(0,\increment){\numbdashes}%
    {\vrule width .4pt height \lengthdash\unitlength}
\arrowtype=#4 \ypos=#2
\ifnum\arrowtype<0 \advance\arrowtype by 7 \fi
\ifcase\arrowtype
\or \advance\ypos by \arrowlength \advance\ypos by -40
    \put(#1,\ypos){\vector(0,1){\lengthdash}}
    \advance\ypos by -40
    \put(#1,\ypos){\vector(0,1){\lengthdash}}
\or \advance\ypos by 10
    \put(#1,\ypos){\vector(0,1){\lengthdash}}
    \advance\ypos by \arrowlength \advance\ypos by -40
    \put(#1,\ypos){\vector(0,1){\lengthdash}}
\or \advance\ypos by \arrowlength \advance\ypos by -40
    \put(#1,\ypos){\vector(0,1){\lengthdash}}
\or \advance\ypos by 10
    \put(#1,\ypos){\vector(0,-1){\lengthdash}}
\or \advance\ypos by 10
    \put(#1,\ypos){\vector(0,-1){\lengthdash}}
    \advance\ypos by \arrowlength \advance\ypos by -40
    \put(#1,\ypos){\vector(0,-1){\lengthdash}}
\or \advance\ypos by 10
    \put(#1,\ypos){\vector(0,-1){\lengthdash}}
    \advance\ypos by 40
    \put(#1,\ypos){\vector(0,-1){\lengthdash}}
\fi
}}
 
\def\puthmorphism(#1,#2)[#3`#4`#5]#6#7#8{{%
\xpos #1
\ypos #2
\width #6
\arrowlength #6
\arrowtype=#7
\putbox(\xpos,\ypos){#3\vphantom{#4}}%
{\advance \xpos by\arrowlength
\putbox(\xpos,\ypos){\vphantom{#3}#4}}%
\horsize{\tempcounta}{#3}%
\horsize{\tempcountb}{#4}%
\divide \tempcounta by2
\divide \tempcountb by2
\advance \tempcounta by30
\advance \tempcountb by30
\advance \xpos by\tempcounta
\advance \arrowlength by-\tempcounta
\advance \arrowlength by-\tempcountb
\putvector(\xpos,\ypos)(1,0)\arrowlength\arrowtype
\divide \arrowlength by2
\advance \xpos by\arrowlength
\vertsize{\tempcounta}{#5}%
\divide\tempcounta by2
\advance \tempcounta by20
\if a#8 %
   \advance \ypos by\tempcounta
   \putbox(\xpos,\ypos){#5}%
\else
   \advance \ypos by-\tempcounta
   \putbox(\xpos,\ypos){#5}%
\fi}}
 
\def\putvmorphism(#1,#2)[#3`#4`#5]#6#7#8{{%
\xpos #1
\ypos #2
\arrowlength #6
\arrowtype #7
\settowidth{\xlen}{$#5$}%
\putbox(\xpos,\ypos){#3}%
{\advance \ypos by-\arrowlength
\putbox(\xpos,\ypos){#4}}%
{\advance\arrowlength by-140
\advance \ypos by-70
\ifdim\xlen>0pt
   \if m#8%
      \putsplitvector(\xpos,\ypos)\arrowlength\arrowtype
   \else
   \putvector(\xpos,\ypos)(0,-1)\arrowlength\arrowtype
   \fi
\else
   \putvector(\xpos,\ypos)(0,-1)\arrowlength\arrowtype
\fi}%
\ifdim\xlen>0pt
   \divide \arrowlength by2
   \advance\ypos by-\arrowlength
   \if l#8%
      \advance \xpos by-40
      \putrbox(\xpos,\ypos){#5}%
   \else\if r#8%
      \advance \xpos by40
      \putlbox(\xpos,\ypos){#5}%
   \else
      \putbox(\xpos,\ypos){#5}%
   \fi\fi
\fi
}}
 
\def\putsquarep<#1>(#2)[#3;#4`#5`#6`#7]{{%
\setsqparms[#1]%
\setpos(#2)%
\settokens`#3`%
\puthmorphism(\xpos,\ypos)[\tokenc`\tokend`{#7}]{\width}{\arrowtyped}b%
\advance\ypos by \height
\puthmorphism(\xpos,\ypos)[\tokena`\tokenb`{#4}]{\width}{\arrowtypea}a%
\putvmorphism(\xpos,\ypos)[``{#5}]{\height}{\arrowtypeb}l%
\advance\xpos by \width
\putvmorphism(\xpos,\ypos)[``{#6}]{\height}{\arrowtypec}r%
}}
 
\def\putsquare{\@ifnextchar <{\putsquarep}{\putsquarep%
   <\arrowtypea`\arrowtypeb`\arrowtypec`\arrowtyped;\width`\height>}}
\def\square{\@ifnextchar< {\squarep}{\squarep
   <\arrowtypea`\arrowtypeb`\arrowtypec`\arrowtyped;\width`\height>}}
\def\squarep<#1>[#2`#3`#4`#5;#6`#7`#8`#9]{{
\setsqparms[#1]
\diagram
\putsquarep<\arrowtypea`\arrowtypeb`\arrowtypec`
\arrowtyped;\width`\height>
(0,0)[#2`#3`#4`{#5};#6`#7`#8`{#9}]
\enddiagram
}}                                                 
\def\putptrianglep<#1>(#2,#3)[#4`#5`#6;#7`#8`#9]{{%
\settriparms[#1]%
\xpos=#2 \ypos=#3
\advance\ypos by \height
\puthmorphism(\xpos,\ypos)[#4`#5`{#7}]{\height}{\arrowtypea}a%
\putvmorphism(\xpos,\ypos)[`#6`{#8}]{\height}{\arrowtypeb}l%
\advance\xpos by\height
\putmorphism(\xpos,\ypos)(-1,-1)[``{#9}]{\height}{\arrowtypec}r%
}}
 
\def\putptriangle{\@ifnextchar <{\putptrianglep}{\putptrianglep
   <\arrowtypea`\arrowtypeb`\arrowtypec;\height>}}
\def\ptriangle{\@ifnextchar <{\ptrianglep}{\ptrianglep
   <\arrowtypea`\arrowtypeb`\arrowtypec;\height>}}
\def\ptrianglep<#1>[#2`#3`#4;#5`#6`#7]{{
\settriparms[#1]
\diagram
\putptrianglep<\arrowtypea`\arrowtypeb`
\arrowtypec;\height>
(0,0)[#2`#3`#4;#5`#6`{#7}]
\enddiagram
}}                                            
 
\def\putqtrianglep<#1>(#2,#3)[#4`#5`#6;#7`#8`#9]{{%
\settriparms[#1]%
\xpos=#2 \ypos=#3
\advance\ypos by\height
\puthmorphism(\xpos,\ypos)[#4`#5`{#7}]{\height}{\arrowtypea}a%
\putmorphism(\xpos,\ypos)(1,-1)[``{#8}]{\height}{\arrowtypeb}l%
\advance\xpos by\height
\putvmorphism(\xpos,\ypos)[`#6`{#9}]{\height}{\arrowtypec}r%
}}
 
\def\putqtriangle{\@ifnextchar <{\putqtrianglep}{\putqtrianglep
   <\arrowtypea`\arrowtypeb`\arrowtypec;\height>}}
\def\qtriangle{\@ifnextchar <{\qtrianglep}{\qtrianglep
   <\arrowtypea`\arrowtypeb`\arrowtypec;\height>}}
\def\qtrianglep<#1>[#2`#3`#4;#5`#6`#7]{{
\settriparms[#1]
\width=\height                                
\diagram
\putqtrianglep<\arrowtypea`\arrowtypeb`
\arrowtypec;\height>
(0,0)[#2`#3`#4;#5`#6`{#7}]
\enddiagram
}}
 
\def\putdtrianglep<#1>(#2,#3)[#4`#5`#6;#7`#8`#9]{{%
\settriparms[#1]%
\xpos=#2 \ypos=#3
\puthmorphism(\xpos,\ypos)[#5`#6`{#9}]{\height}{\arrowtypec}b%
\advance\xpos by \height \advance\ypos by\height
\putmorphism(\xpos,\ypos)(-1,-1)[``{#7}]{\height}{\arrowtypea}l%
\putvmorphism(\xpos,\ypos)[#4``{#8}]{\height}{\arrowtypeb}r%
}}
 
\def\putdtriangle{\@ifnextchar <{\putdtrianglep}{\putdtrianglep
   <\arrowtypea`\arrowtypeb`\arrowtypec;\height>}}
\def\dtriangle{\@ifnextchar <{\dtrianglep}{\dtrianglep
   <\arrowtypea`\arrowtypeb`\arrowtypec;\height>}}
\def\dtrianglep<#1>[#2`#3`#4;#5`#6`#7]{{
\settriparms[#1]
\width=\height                                
\diagram
\putdtrianglep<\arrowtypea`\arrowtypeb`
\arrowtypec;\height>
(0,0)[#2`#3`#4;#5`#6`{#7}]
\enddiagram
}}
 
\def\putbtrianglep<#1>(#2,#3)[#4`#5`#6;#7`#8`#9]{{%
\settriparms[#1]%
\xpos=#2 \ypos=#3
\puthmorphism(\xpos,\ypos)[#5`#6`{#9}]{\height}{\arrowtypec}b%
\advance\ypos by\height
\putmorphism(\xpos,\ypos)(1,-1)[``{#8}]{\height}{\arrowtypeb}r%
\putvmorphism(\xpos,\ypos)[#4``{#7}]{\height}{\arrowtypea}l%
}}
 
\def\putbtriangle{\@ifnextchar <{\putbtrianglep}{\putbtrianglep
   <\arrowtypea`\arrowtypeb`\arrowtypec;\height>}}
\def\btriangle{\@ifnextchar <{\btrianglep}{\btrianglep
   <\arrowtypea`\arrowtypeb`\arrowtypec;\height>}}
\def\btrianglep<#1>[#2`#3`#4;#5`#6`#7]{{
\settriparms[#1]
\width=\height                               
\diagram
\putbtrianglep<\arrowtypea`\arrowtypeb`
\arrowtypec;\height>
(0,0)[#2`#3`#4;#5`#6`{#7}]
\enddiagram
}}
 
\def\putAtrianglep<#1>(#2,#3)[#4`#5`#6;#7`#8`#9]{{%
\settriparms[#1]%
\xpos=#2 \ypos=#3
{\multiply \height by2
\puthmorphism(\xpos,\ypos)[#5`#6`{#9}]{\height}{\arrowtypec}b}%
\advance\xpos by\height \advance\ypos by\height
\putmorphism(\xpos,\ypos)(-1,-1)[#4``{#7}]{\height}{\arrowtypea}l%
\putmorphism(\xpos,\ypos)(1,-1)[``{#8}]{\height}{\arrowtypeb}r%
}}
 
\def\putAtriangle{\@ifnextchar <{\putAtrianglep}{\putAtrianglep
   <\arrowtypea`\arrowtypeb`\arrowtypec;\height>}}
\def\Atriangle{\@ifnextchar <{\Atrianglep}{\Atrianglep
   <\arrowtypea`\arrowtypeb`\arrowtypec;\height>}}
\def\Atrianglep<#1>[#2`#3`#4;#5`#6`#7]{{
\settriparms[#1]
\width=\height                                     
\diagram
\putAtrianglep<\arrowtypea`\arrowtypeb`
\arrowtypec;\height>
(0,0)[#2`#3`#4;#5`#6`{#7}]
\enddiagram
}}
 
\def\putAtrianglepairp<#1>(#2)[#3;#4`#5`#6`#7`#8]{{%
\settripairparms[#1]%
\setpos(#2)%
\settokens`#3`%
\puthmorphism(\xpos,\ypos)[\tokenb`\tokenc`{#7}]{\height}{\arrowtyped}b%
\advance\xpos by\height
\puthmorphism(\xpos,\ypos)[\phantom{\tokenc}`\tokend`{#8}]%
{\height}{\arrowtypee}b%
\advance\ypos by\height
\putmorphism(\xpos,\ypos)(-1,-1)[\tokena``{#4}]{\height}{\arrowtypea}l%
\putvmorphism(\xpos,\ypos)[``{#5}]{\height}{\arrowtypeb}m%
\putmorphism(\xpos,\ypos)(1,-1)[``{#6}]{\height}{\arrowtypec}r%
}}
 
\def\putAtrianglepair{\@ifnextchar <{\putAtrianglepairp}{\putAtrianglepairp%
   <\arrowtypea`\arrowtypeb`\arrowtypec`\arrowtyped`\arrowtypee;\height>}}
\def\Atrianglepair{\@ifnextchar <{\Atrianglepairp}{\Atrianglepairp%
   <\arrowtypea`\arrowtypeb`\arrowtypec`\arrowtyped`\arrowtypee;\height>}}
 
\def\Atrianglepairp<#1>[#2;#3`#4`#5`#6`#7]{{
\settripairparms[#1]
\settokens`#2`
\width=\height                                
\diagram
\putAtrianglepairp                            
<\arrowtypea`\arrowtypeb`\arrowtypec`
\arrowtyped`\arrowtypee;\height>
(0,0)[{#2};#3`#4`#5`#6`{#7}]
\enddiagram
}}
 
\def\putVtrianglep<#1>(#2,#3)[#4`#5`#6;#7`#8`#9]{{%
\settriparms[#1]%
\xpos=#2 \ypos=#3
\advance\ypos by\height
{\multiply\height by2
\puthmorphism(\xpos,\ypos)[#4`#5`{#7}]{\height}{\arrowtypea}a}%
\putmorphism(\xpos,\ypos)(1,-1)[`#6`{#8}]{\height}{\arrowtypeb}l%
\advance\xpos by\height
\advance\xpos by\height
\putmorphism(\xpos,\ypos)(-1,-1)[``{#9}]{\height}{\arrowtypec}r%
}}
 
\def\putVtriangle{\@ifnextchar <{\putVtrianglep}{\putVtrianglep
   <\arrowtypea`\arrowtypeb`\arrowtypec;\height>}}
\def\Vtriangle{\@ifnextchar <{\Vtrianglep}{\Vtrianglep
   <\arrowtypea`\arrowtypeb`\arrowtypec;\height>}}
\def\Vtrianglep<#1>[#2`#3`#4;#5`#6`#7]{{
\settriparms[#1]
\width=\height                                 
\diagram
\putVtrianglep<\arrowtypea`\arrowtypeb`
\arrowtypec;\height>
(0,0)[#2`#3`#4;#5`#6`{#7}]
\enddiagram
}}
 
\def\putVtrianglepairp<#1>(#2)[#3;#4`#5`#6`#7`#8]{{
\settripairparms[#1]%
\setpos(#2)%
\settokens`#3`%
\advance\ypos by\height
\putmorphism(\xpos,\ypos)(1,-1)[`\tokend`{#6}]{\height}{\arrowtypec}l%
\puthmorphism(\xpos,\ypos)[\tokena`\tokenb`{#4}]{\height}{\arrowtypea}a%
\advance\xpos by\height
\puthmorphism(\xpos,\ypos)[\phantom{\tokenb}`\tokenc`{#5}]%
{\height}{\arrowtypeb}a%
\putvmorphism(\xpos,\ypos)[``{#7}]{\height}{\arrowtyped}m%
\advance\xpos by\height
\putmorphism(\xpos,\ypos)(-1,-1)[``{#8}]{\height}{\arrowtypee}r%
}}
 
\def\putVtrianglepair{\@ifnextchar <{\putVtrianglepairp}{\putVtrianglepairp%
    <\arrowtypea`\arrowtypeb`\arrowtypec`\arrowtyped`\arrowtypee;\height>}}
\def\Vtrianglepair{\@ifnextchar <{\Vtrianglepairp}{\Vtrianglepairp%
    <\arrowtypea`\arrowtypeb`\arrowtypec`\arrowtyped`\arrowtypee;\height>}}
\def\Vtrianglepairp<#1>[#2;#3`#4`#5`#6`#7]{{
\settripairparms[#1]
\settokens`#2`
\diagram
\putVtrianglepairp                             
<\arrowtypea`\arrowtypeb`\arrowtypec`
\arrowtyped`\arrowtypee;\height>
(0,0)[{#2};#3`#4`#5`#6`{#7}]
\enddiagram
}}

\def\putCtrianglep<#1>(#2,#3)[#4`#5`#6;#7`#8`#9]{{%
\settriparms[#1]%
\xpos=#2 \ypos=#3
\advance\ypos by\height
\putmorphism(\xpos,\ypos)(1,-1)[``{#9}]{\height}{\arrowtypec}l%
\advance\xpos by\height
\advance\ypos by\height
\putmorphism(\xpos,\ypos)(-1,-1)[#4`#5`{#7}]{\height}{\arrowtypea}l%
{\multiply\height by 2
\putvmorphism(\xpos,\ypos)[`#6`{#8}]{\height}{\arrowtypeb}r}%
}}
 
\def\putCtriangle{\@ifnextchar <{\putCtrianglep}{\putCtrianglep
    <\arrowtypea`\arrowtypeb`\arrowtypec;\height>}}
\def\Ctriangle{\@ifnextchar <{\Ctrianglep}{\Ctrianglep
    <\arrowtypea`\arrowtypeb`\arrowtypec;\height>}}
\def\Ctrianglep<#1>[#2`#3`#4;#5`#6`#7]{{
\settriparms[#1]
\width=\height                               
\diagram
\putCtrianglep<\arrowtypea`\arrowtypeb`
\arrowtypec;\height>
(0,0)[#2`#3`#4;#5`#6`{#7}]
\enddiagram
}}                                           
\def\putDtrianglep<#1>(#2,#3)[#4`#5`#6;#7`#8`#9]{{%
\settriparms[#1]%
\xpos=#2 \ypos=#3
\advance\xpos by\height \advance\ypos by\height
\putmorphism(\xpos,\ypos)(-1,-1)[``{#9}]{\height}{\arrowtypec}r%
\advance\xpos by-\height \advance\ypos by\height
\putmorphism(\xpos,\ypos)(1,-1)[`#5`{#8}]{\height}{\arrowtypeb}r%
{\multiply\height by 2
\putvmorphism(\xpos,\ypos)[#4`#6`{#7}]{\height}{\arrowtypea}l}%
}}
 
\def\putDtriangle{\@ifnextchar <{\putDtrianglep}{\putDtrianglep
    <\arrowtypea`\arrowtypeb`\arrowtypec;\height>}}
\def\Dtriangle{\@ifnextchar <{\Dtrianglep}{\Dtrianglep
   <\arrowtypea`\arrowtypeb`\arrowtypec;\height>}}
\def\Dtrianglep<#1>[#2`#3`#4;#5`#6`#7]{{
\settriparms[#1]
\width=\height                              
\diagram
\putDtrianglep<\arrowtypea`\arrowtypeb`
\arrowtypec;\height>
(0,0)[#2`#3`#4;#5`#6`{#7}]
\enddiagram
}}                                          
\def\setrecparms[#1`#2]{\width=#1 \height=#2}%
 
\def\recursep<#1`#2>[#3;#4`#5`#6`#7`#8]{{\m@th
\width=#1 \height=#2
\settokens`#3`
\settowidth{\tempdimen}{$\tokena$}
\ifdim\tempdimen=0pt
  \savebox{\tempboxa}{\hbox{$\tokenb$}}%
  \savebox{\tempboxb}{\hbox{$\tokend$}}%
  \savebox{\tempboxc}{\hbox{$#6$}}%
\else
  \savebox{\tempboxa}{\hbox{$\hbox{$\tokena$}\times\hbox{$\tokenb$}$}}%
  \savebox{\tempboxb}{\hbox{$\hbox{$\tokena$}\times\hbox{$\tokend$}$}}%
  \savebox{\tempboxc}{\hbox{$\hbox{$\tokena$}\times\hbox{$#6$}$}}%
\fi
\ypos=\height
\divide\ypos by 2
\xpos=\ypos
\advance\xpos by \width
\bfig
\putCtrianglep<-1`1`1;\ypos>(0,0)[`\tokenc`;#5`#6`{#7}]%
\puthmorphism(\ypos,0)[\tokend`\usebox{\tempboxb}`{#8}]{\width}{-1}b%
\puthmorphism(\ypos,\height)[\tokenb`\usebox{\tempboxa}`{#4}]{\width}{-1}a%
\advance\ypos by \width
\putvmorphism(\ypos,\height)[``\usebox{\tempboxc}]{\height}1r%
\efig
}}
 
\def\recurse{\@ifnextchar <{\recursep}{\recursep<\width`\height>}}
 
\def\puttwohmorphisms(#1,#2)[#3`#4;#5`#6]#7#8#9{{%
\puthmorphism(#1,#2)[#3`#4`]{#7}0a
\ypos=#2
\advance\ypos by 20
\puthmorphism(#1,\ypos)[\phantom{#3}`\phantom{#4}`#5]{#7}{#8}a
\advance\ypos by -40
\puthmorphism(#1,\ypos)[\phantom{#3}`\phantom{#4}`#6]{#7}{#9}b
}}
 
\def\puttwovmorphisms(#1,#2)[#3`#4;#5`#6]#7#8#9{{%
\putvmorphism(#1,#2)[#3`#4`]{#7}0a
\xpos=#1
\advance\xpos by -20
\putvmorphism(\xpos,#2)[\phantom{#3}`\phantom{#4}`#5]{#7}{#8}l
\advance\xpos by 40
\putvmorphism(\xpos,#2)[\phantom{#3}`\phantom{#4}`#6]{#7}{#9}r
}}
 
\def\puthcoequalizer(#1)[#2`#3`#4;#5`#6`#7]#8#9{{%
\setpos(#1)%
\puttwohmorphisms(\xpos,\ypos)[#2`#3;#5`#6]{#8}11%
\advance\xpos by #8
\puthmorphism(\xpos,\ypos)[\phantom{#3}`#4`#7]{#8}1{#9}
}}
 
\def\putvcoequalizer(#1)[#2`#3`#4;#5`#6`#7]#8#9{{%
\setpos(#1)%
\puttwovmorphisms(\xpos,\ypos)[#2`#3;#5`#6]{#8}11%
\advance\ypos by -#8
\putvmorphism(\xpos,\ypos)[\phantom{#3}`#4`#7]{#8}1{#9}
}}
 
\def\putthreehmorphisms(#1)[#2`#3;#4`#5`#6]#7(#8)#9{{%
\setpos(#1) \settypes(#8)
\if a#9 %
     \vertsize{\tempcounta}{#5}%
     \vertsize{\tempcountb}{#6}%
     \ifnum \tempcounta<\tempcountb \tempcounta=\tempcountb \fi
\else
     \vertsize{\tempcounta}{#4}%
     \vertsize{\tempcountb}{#5}%
     \ifnum \tempcounta<\tempcountb \tempcounta=\tempcountb \fi
\fi
\advance \tempcounta by 60
\puthmorphism(\xpos,\ypos)[#2`#3`#5]{#7}{\arrowtypeb}{#9}
\advance\ypos by \tempcounta
\puthmorphism(\xpos,\ypos)[\phantom{#2}`\phantom{#3}`#4]{#7}{\arrowtypea}{#9}
\advance\ypos by -\tempcounta \advance\ypos by -\tempcounta
\puthmorphism(\xpos,\ypos)[\phantom{#2}`\phantom{#3}`#6]{#7}{\arrowtypec}{#9}
}}
 
\def\setarrowtoks[#1`#2`#3`#4`#5`#6]{%
\def\toka{#1}
\def\tokb{#2}
\def\tokc{#3}
\def\tokd{#4}
\def\toke{#5}
\def\tokf{#6}
}
\def\hex{\@ifnextchar <{\hexp}{\hexp<1000`400>}}
\def\hexp<#1`#2>[#3`#4`#5`#6`#7`#8;#9]{%
\setarrowtoks[#9]
\yext=#2 \advance \yext by #2
\xext=#1 \advance\xext by \yext
\bfig
\putCtriangle<-1`0`1;#2>(0,0)[`#5`;\tokb``\tokd]
\xext=#1 \yext=#2 \advance \yext by #2
\putsquare<1`0`0`1;\xext`\yext>(#2,0)[#3`#4`#7`#8;\toka```\tokf]
\advance \xext by #2
\putDtriangle<0`1`-1;#2>(\xext,0)[`#6`;`\tokc`\toke]
\efig
}
\makeatother